\documentclass[aps,pre,reprint,longbibliography,floatfix]{revtex4-2}

\usepackage{amsmath,amssymb,amsfonts,mathtools,bm}
\usepackage{graphicx}
\usepackage{microtype}
\usepackage{xcolor}
\usepackage{hyperref}

\hypersetup{colorlinks=true,linkcolor=blue!50!black,citecolor=blue!50!black,urlcolor=blue!50!black}

\newcommand{\dd}{\mathrm{d}}
\newcommand{\ee}{\mathrm{e}}
\newcommand{\avg}[1]{\left\langle #1 \right\rangle}
\newcommand{\Var}{\mathrm{Var}}
\newcommand{\bud}{\mathrm{bud}}
\newcommand{\shear}{\mathrm{shear}}
\newcommand{\grow}{\mathrm{grow}}
\newcommand{\ext}{\mathrm{ext}}
\newcommand{\iso}{\mathrm{iso}}
\newcommand{\nr}{\mathrm{nr}}
\newcommand{\divv}{\mathrm{div}}
\newcommand{\runin}[1]{\smallskip\noindent\textit{#1.—}}
\newcommand{\placeholdergraphic}[1]{%
  \fbox{\begin{minipage}[c][0.24\textheight][c]{0.9\linewidth}
  \centering Missing figure\\[2pt]\texttt{\detokenize{#1}}
  \end{minipage}}}
\newcommand{\smartincludegraphics}[2][]{%
  \IfFileExists{#2}{\includegraphics[#1]{#2}}{%
    \IfFileExists{#2.pdf}{\includegraphics[#1]{#2.pdf}}{%
      \IfFileExists{#2.png}{\includegraphics[#1]{#2.png}}{%
        \IfFileExists{#2.jpg}{\includegraphics[#1]{#2.jpg}}{%
          \IfFileExists{#2.jpeg}{\includegraphics[#1]{#2.jpeg}}{%
            \placeholdergraphic{#2}}}}}}}

\begin{document}

\title{Theory of post-jamming rigidity in feedback-regulated cellular packings}
\author{Pawe{\l} Gniewek}
\affiliation{Independent Researcher}
\email{gniewko.pablo@gmail.com}
\makeatletter\let\@date\@empty\makeatother

\begin{abstract}
Budding-cell packings jam before all buds are mechanically constrained, so the post-jamming state is set not by the pressure $P$ alone but also by the fraction $u$ of buds that remain unconstrained. We develop a mean-field theory in these two variables for this regime. Stress feedback suppresses growth on the loaded buds, so continued growth is redirected onto the remaining free ones. As those buds are completed they add contacts that raise the excess coordination without a comparable rise in prestress. We introduce a modified Maxwell count, a bud-depletion relation, and a flux-partition argument to predict the post-jamming coordination, the density at which the reservoir of initially free buds is exhausted, and how strong feedback can stiffen the packing while generating little internal pressure. Because the added contacts raise the rigidity while the prestress stays low, we conclude that feedback-regulated growth provides a distinct mechanism of self-rigidification.
\end{abstract}

\maketitle

Confined microbial populations can jam under their own growth \cite{Delarue2016}. Growth-induced mechanical stress has been recognized as an innate feature of spatially confined cell populations \cite{Delarue2025}. In budding yeast, experiments showed that the contact pressures generated in this way feed back on cell growth \cite{Delarue2016}, and simulations showed that the same feedback can raise the contact number and strongly stiffen the packing while adding little pressure during further growth \cite{Gniewek2019}. Ordinary compression-driven jamming of purely repulsive particles does not have this feature \cite{OHern2003,LiuNagel2010}. 
More broadly, proliferating active matter---in which replication injects both biomass and new degrees of freedom---has emerged as a distinct class of driven systems with no counterpart in fixed-particle-number active matter \cite{Hallatschek2023}.
Jamming transitions also arise in biological settings, from the fluid-to-solid transition that sculpts the vertebrate body axis~\cite{Mongera2018} to unjamming in tissue development~\cite{Atia2021}. In budding-cell populations, however, jamming occurs while a finite fraction $u_J$ of buds is still mechanically unconstrained, which makes the post-jamming regime distinct from ordinary compression-driven jamming. Those unconstrained buds leave residual growth modes at the jamming point, so pressure alone does not specify the post-jamming state. Here we develop a mean-field theory to quantitatively characterize this regime. Using the fraction of unconstrained buds $u$ as the additional state variable, the model predicts how that population is depleted above jamming, which then explains why strong feedback can rigidify the packing without building up much internal pressure. This establishes growth as a separate route to rigidity, distinguishing it from the compression axis of ordinary jamming. We show that this separation is sharp---the packing's stiffness approaches a finite plateau while the pressure vanishes exponentially with feedback strength. Finally, the analysis of the vibrational spectrum places the feedback-rigidified packings within the jamming universality class once the unconstrained-bud modes are properly accounted for.

\begin{figure}[t]
\centering
\smartincludegraphics[width=0.95\columnwidth]{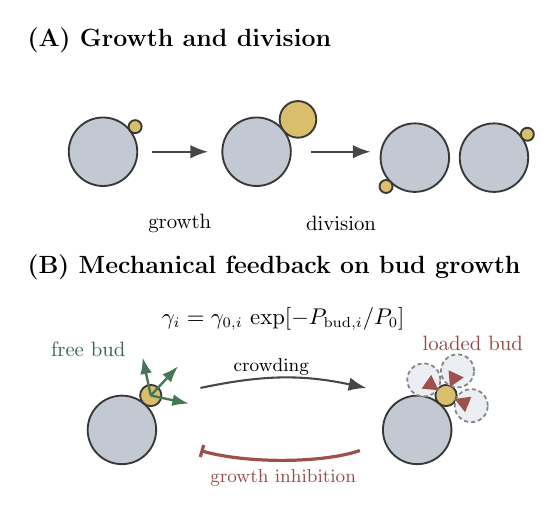}
\caption{
Budding-cell model.
(\textbf{A}) A cell grows by bud expansion until mother--daughter separation creates two newborn cells.
(\textbf{B}) Local feedback law, Eq.~\eqref{eq:growthlaw}. An unconstrained bud grows near its innate rate $\gamma_{0,i}$, whereas crowding by neighboring lobes loads the bud and suppresses growth through the local pressure $P_{\bud,i}$.
}
\label{fig:model}
\end{figure}

\runin{Model} We use the two-dimensional budding-cell model of Ref.~\cite{Gniewek2019}. Each cell consists of a rigid mother lobe and a growing bud [Fig.~\ref{fig:model}(\textbf{A})]. Overlapping lobes interact repulsively, and after each small growth step the packing relaxes quasistatically. The local growth law is
\begin{equation}
\gamma_i=\gamma_{0,i}\cdot \ee^{-P_{\bud,i}/{P_0}}
\label{eq:growthlaw}
\end{equation}
where $\gamma_{0,i}$ is the innate zero-pressure growth rate assigned at cell birth, $P_{\bud,i}$ is the pressure on bud $i$, and $P_0$ is the feedback pressure scale [Fig.~\ref{fig:model}(\textbf{B})]. Smaller values of $P_0$ correspond to stronger feedback. Simulation details are given in Appendix~\ref{app:protocol}.

Let $N_f$ be the number of rattlers, let $N_u$ be the number of unconstrained buds among the remaining cells, and let $N_{\nr}=N-N_f$ be the number of nonrattler cells. We define
\begin{equation}
u\equiv\frac{N_u}{N_{\nr}},
\qquad
Z\equiv\frac{N_c}{N_{\nr}},
\label{eq:intensives}
\end{equation}
where $N_c$ counts contacts twice. The post-jamming theory uses pressure $P$ and the unconstrained-bud fraction $u$ as its state variables. It applies when jamming leaves a finite population of unconstrained buds, $u_J>0$. If instead $u_J=0$, feedback may still slow growth, but there is no residual underconstrained sector to deplete, so the specific post-jamming structural mechanism discussed here is absent.

\runin{Modified counting} A finite population of unconstrained buds modifies the standard central-force isostatic count~\cite{vanHecke2010}. In doubled-contact counting, the finite-size isostatic relation is
\begin{equation}
N_c^{\iso}=6N_{\nr}-2N_u-2.
\label{eq:codeiso}
\end{equation}
Note that removing the two global translations would by itself give $6N_{\nr}-2N_u-4$. A periodic packing with only that many counted contacts is marginal but cannot support a finite pressure or have a positive bulk modulus $B>0$. Therefore, one additional physical contact is needed to sustain compression, just as is observed in finite-size jamming of sphere packings \cite{Goodrich2012}. In doubled-contact counting that extra constraint contributes $+2$ to the count, which yields Eq.~\eqref{eq:codeiso}. A finite-size derivation of Eq.~\eqref{eq:codeiso} and the associated contact-budget argument are given in Appendix~\ref{app:count}. Dividing by $N_{\nr}$ gives
\begin{equation}
Z_{\iso}(u)=6-2u-\frac{2}{N_{\nr}}\simeq 6-2u.
\label{eq:Ziso}
\end{equation}
This relation is analogous to the reduced isostatic coordination found in other nonspherical packings \cite{Mailman2009,Schreck2010}. Studies of dimer and other composite-particle packings confirm that proper identification of unconstrained internal modes---rotational rattlers in dimer packings play an analogous role to the unconstrained buds here---restores isostaticity at jamming \cite{VanderWerf2018,Ikeda2021}. Once these internal modes are counted, such packings fall within the standard jamming universality class~\cite{Brito2018}. The relevant distance from marginality is therefore
\begin{equation}
\Delta Z\equiv Z-Z_{\iso}(u).
\label{eq:DeltaZdef}
\end{equation}

\begin{figure*}[t]
\centering
\smartincludegraphics[width=0.85\textwidth]{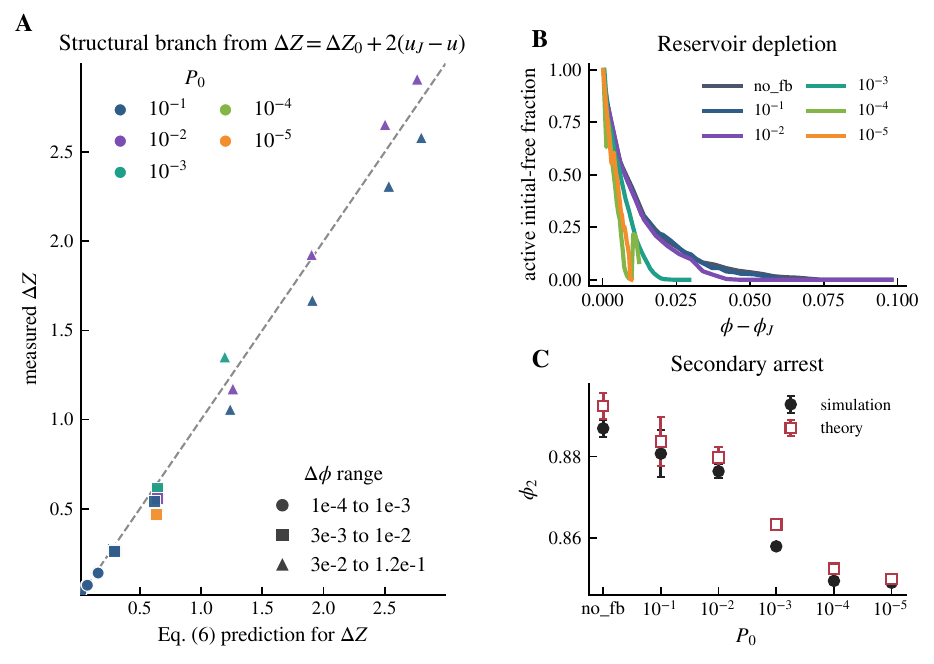}
\caption{
Structure for system size $L=15D_0$, where $D_0$ is a characteristic cell size.
(\textbf{A}) Measured excess coordination $\Delta Z$ versus the counting prediction $\Delta Z_0+2(u_J-u)$, tested over $(P_0,\Delta\phi)$.
(\textbf{B}) Fraction of the buds that were unconstrained at $\phi_J$ and remain unconstrained at higher density, plotted against $\phi-\phi_J$.
(\textbf{C}) Depletion density $\phi_2(P_0)$ from the simulations and from Eq.~\eqref{eq:phi2general}. Error bars are standard errors (SE) of the mean across runs for each $P_0$. 
}
\label{fig:structure}
\end{figure*}

Each time an unconstrained bud is completed it forms two physical contacts (four in doubled counting), so $Z$ increases by $4/N_{\nr}$, whereas the modified isostatic count increases only by $2/N_{\nr}$ (one physical contact). One of the two contacts removes the bud's free rotation, while the second is a surplus contact that adds one state of self-stress. The excess coordination therefore grows by $2/N_{\nr}$ per completion. Hence
\begin{equation}
\Delta Z(P,u)=\Delta Z_0(P)+2(u_J-u),
\label{eq:DZtotal}
\end{equation}
and
\begin{equation}
Z(P,u)=6+2u_J-4u+\Delta Z_0(P).
\label{eq:ZofPu}
\end{equation} 
Thus new contacts are created by two mechanisms: increased compaction (as in athermal granular packings) and biomass-driven depletion of buds that were previously unconstrained. Fig.~\ref{fig:structure}(\textbf{A}) tests this relation over a broad set of $(P_0,\Delta\phi)$ values. It holds over three orders of magnitude in the system compaction $\Delta\phi$ and more than four orders of magnitude in the feedback strength. Small deviations are expected and can be attributed to finite-size effects that decay as $\sim N^{-1}_{\nr}$, to one-contact intermediate states not resolved in the reduced counting argument, or, at the largest $\Delta\phi$, to cell divisions that begin to repopulate the unconstrained sector. The value of $u_J$ in Eq.~\eqref{eq:ZofPu} can be predicted with a mean-field counting argument (Appendix~\ref{app:count}) yielding a parameter-free estimate $u_J\approx0.20$, which is close to the numerically measured average $u_J\simeq0.23$. The finite-size effects are discussed in Appendix~\ref{app:sizeeffects}, and the injection of new degrees of freedom is discussed in Appendix~\ref{app:depletion}.

\runin{Bud depletion} To follow the free-bud reservoir above jamming, we label the buds that are unconstrained at $\phi_J$. For each such bud $\alpha$, let $a_\alpha^\ast$ be the added bud area required for mechanical completion. We define
\begin{equation}
H(a)=\Pr(a^\ast\le a),
\qquad
S(a)=1-H(a),
\label{eq:HS}
\end{equation}
where $S(a)$ is the survival function of these completion thresholds. In the interval before new divisions matter, the surviving unconstrained-bud fraction is
\begin{equation}
u(a)=u_J S(a).
\label{eq:uofa}
\end{equation}
Using the Kaplan--Meier construction, we estimate the survival function from numerical data at varying feedback strength and compaction; see Fig.~\ref{fig:structure}(\textbf{B}). For populations with no feedback or very weak feedback, growth occurs in both the compressed and free sectors, and free buds become constrained over a broad range of compactions. That changes as stress feedback on bud growth becomes stronger. In these cases we observe a sharper drop in the population of unconstrained buds, and the full bud mechanical completion occurs at lower densities than for weak feedback.

To analyze this quantitatively, let
\begin{equation}
\chi_c(a;P_0)\equiv \avg{\ee^{-P_{\bud}/P_0}}_c
\label{eq:chi}
\end{equation}
be the mean growth factor of the compressed buds. Approximating the nonrattler density by its jamming value $n_J$ over this narrow interval, the density increment obeys
\begin{equation}
\frac{\dd\phi}{\dd a}=n_J\bigl[u(a)+(1-u(a))\chi_c(a;P_0)\bigr].
\label{eq:dphida_exact}
\end{equation}
If $\chi_c$ varies only weakly during depletion, we replace it by its interval average $\bar{\chi}(P_0)$ and obtain
\begin{equation}
\begin{split}
\phi(a;P_0)-\phi_J
= n_J\Bigl[&\bar{\chi}(P_0)a \\
&+ \bigl(1-\bar{\chi}(P_0)\bigr)u_J \int_0^a S(a')\,\dd a'\Bigr].
\end{split}
\label{eq:phiofa}
\end{equation}

We define the depletion density operationally by
\begin{equation}
u(\phi_2,P_0)=\varepsilon,
\label{eq:phi2def}
\end{equation}
with $\varepsilon\ll 1$. In a finite system, taking $\varepsilon$ to be of order $N_{\nr,J}^{-1}$, with $N_{\nr,J}$ the nonrattler population at jamming, makes this operational definition coincide with exhaustion of the last tagged unconstrained bud up to the saved-state resolution. If $u_J S(a_\varepsilon)=\varepsilon$, then
\begin{equation}
\begin{split}
\phi_2(P_0)-\phi_J
= n_J\Bigl[&\bar{\chi}(P_0)a_\varepsilon \\
&+ \bigl(1-\bar{\chi}(P_0)\bigr)u_J \int_0^{a_\varepsilon} S(a)\,\dd a\Bigr].
\end{split}
\label{eq:phi2general}
\end{equation}
Thus stronger feedback moves $\phi_2$ closer to $\phi_J$---the same degree of bud completion is reached after a smaller increase in packing fraction because a smaller fraction of growth is spent on the compressed sector. This somewhat surprising trend is observed in the numerical data shown in Fig.~\ref{fig:structure}(\textbf{C}). The simulated depletion density---the packing fraction at which the last bud that was free at jamming becomes complete---tracks Eq.~\eqref{eq:phi2general}, with a small systematic overestimate by the theory. In the strong-feedback limit, $P_0\rightarrow 0^+$, the survival curve develops a sharp edge. We therefore approximate $S(a)$ by a step,
\begin{equation}
S(a)\approx \Theta(a_c-a).
\label{eq:sharpassumption}
\end{equation}
This reduces the depletion density to
\begin{equation}
\phi_2(P_0)\approx \phi_J+n_J a_c\bigl[u_J+(1-u_J)\bar{\chi}(P_0)\bigr].
\label{eq:phi2sharp}
\end{equation}
The impact of threshold broadening in Eq.~\eqref{eq:sharpassumption} and the contribution from the injection of new unconstrained buds by cell division are minor in the strong-feedback regime. The full quantitative analysis is presented in Appendix~\ref{app:depletion}.

\begin{figure*}[t!]
\centering
\smartincludegraphics[width=0.85\textwidth]{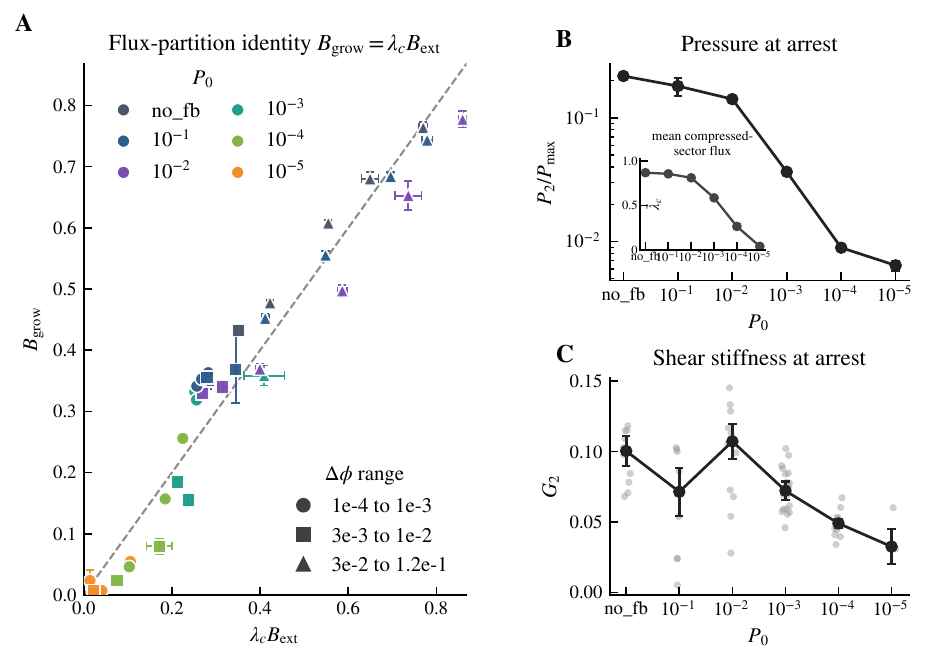}
\caption{
Mechanical consequences of flux partitioning for system size $L=15D_0$.
(\textbf{A}) Measured growth bulk response $B_{\grow}$ versus the prediction $\lambda_c B_{\ext}$ from Eq.~\eqref{eq:BgrowIdentity}, tested over $(P_0,\Delta\phi)$.
(\textbf{B}) Pressure at the depletion point $\phi_2$, normalized by the confluence pressure scale $P_{\max}\approx0.1~\mathrm{a.u.}$ (Appendix~\ref{app:mechanics}). The inset shows the corresponding depletion-interval mean flux fraction $\bar{\lambda}_c(P_0)$. Error bars are SE of the mean across runs.
(\textbf{C}) Shear modulus at the same depletion point, $G_2\equiv G(\phi_2(P_0);P_0)$. Gray points show individual simulations. Error bars are bootstrap SE of the median, using 2000 bootstrap resamples.
}
\label{fig:mechanics}
\end{figure*}

\runin{Mechanical consequences} The same split between free and compressed sectors accounts for the mechanics reported in Ref.~\cite{Gniewek2019}. For a mechanically arrested packing, the mean-field estimates combine an affine bulk response with the standard excess-coordination scaling of the shear response:
\begin{align}
B_{\ext}&=\frac{n_J k\ell_c^2}{8}\,Z,
\label{eq:BextClosed}\\
G&=\frac{n_J k\ell_c^2}{2}M_4\,\Delta Z,
\label{eq:Ggeneral}
\end{align}
where $\ell_c$ is the mean contact length and $M_4\equiv\avg{n_x^2n_y^2}=1/8$ for an isotropic contact network. Thus $B_{\ext}$ is set by the total contact number, while $G$ is set by the excess coordination $\Delta Z$ \cite{Wyart2005EPL,Ellenbroek2006,Goodrich2012}. The derivations of Eqs.~\eqref{eq:BextClosed} and \eqref{eq:Ggeneral} are collected in Appendix~\ref{app:mechanics}.

During a small increment $\dd a$ of bud area, the unconstrained sector contributes $n_Ju\,\dd a$ to $\dd\phi$ and the compressed sector contributes $n_J(1-u)\chi_c\,\dd a$. The fraction of the total density increment entering the compressed sector is therefore
\begin{equation}
\lambda_c(\phi,P_0)=\frac{(1-u)\chi_c}{u+(1-u)\chi_c}.
\label{eq:lambdac}
\end{equation}
In this mean-field description, the pre-existing force network is loaded only by the compressed-sector part of the growth. Hence the growth bulk response is
\begin{equation}
B_{\grow}(\phi,P_0)\equiv \phi\left(\frac{\partial P}{\partial\phi}\right)_{\grow}
=\lambda_c(\phi,P_0)\,B_{\ext}(\phi,P_0).
\label{eq:BgrowIdentity}
\end{equation}
This is the main mechanical prediction of the theory. Fig.~\ref{fig:mechanics}(\textbf{A}) compares this prediction with the measured growth bulk response.
Free buds can expand into voids, but load-bearing buds must push against the contact network.
Feedback controls growth-generated pressure through $\lambda_c$, while rigidity is controlled by the structural term $\Delta Z$.
The two can therefore separate. To see that, note that at the depletion density, the pressure buildup is
\begin{equation}
P_2(P_0)=\int_{\phi_J}^{\phi_2(P_0)}\frac{\dd\phi}{\phi}\,\lambda_c(\phi,P_0)B_{\ext}(\phi,P_0)
\label{eq:P2integral}
\end{equation}
and the packing stiffness is
\begin{equation}
G_2(P_0)=\frac{n_J k\ell_c^2}{2}M_4\Bigl[2(u_J-\varepsilon)+\Delta Z_0\!\bigl(P_2(P_0)\bigr)\Bigr].
\label{eq:G2arrest}
\end{equation}
As $P_0\to 0$, $\lambda_c$ also tends to zero, but the structural term $2(u_J-\varepsilon)$ remains finite. Consequently, even as the prestress at $\phi_2$ diminishes, the stiffness of the packing is maintained. In Fig.~\ref{fig:mechanics}(\textbf{B}), the internal pressure at the depletion point, $P_2$, decreases strongly as the feedback strengthens. Notably, that decrease is closely mirrored by the mean compressed-sector flux fraction, $\bar{\lambda}_c$. 
This suggests that $B_{\rm ext}$ varies only weakly over that compression interval, so $P_2$ decays approximately exponentially as $P_2\propto\exp[-P^\dagger/P_0]$, with $P^\dagger$ a typical bud pressure over the $\phi_2-\phi_J$ interval.
As a result, $P_2$ is indeed driven toward zero, whereas $G_2$ remains finite and is consistent with a low-pressure plateau of order $G_2(0^+)=n_Jk\ell_c^2M_4u_J$ as $P_0 \to 0$ [Fig.~\ref{fig:mechanics}(\textbf{C})]. 
Equivalently, the dimensionless stiffness-to-stress ratio diverges, $G_2/P_2\to\infty$ as $P_0\to0$ (Appendix~\ref{app:mechanics} and Fig.~\ref{figS:mechanicspaired}).
It is worth noting that at the depletion point, the shear modulus $G_2$ is substantial and within an order of magnitude of the internal pressure of the compacted packing at full confluence, i.e., $G_2/P_{\max} \approx 0.2-1$. The estimate of $P_{\max}$ is given in Appendix~\ref{app:mechanics}.

\begin{figure}[t]
\centering
\smartincludegraphics[width=0.97\columnwidth]{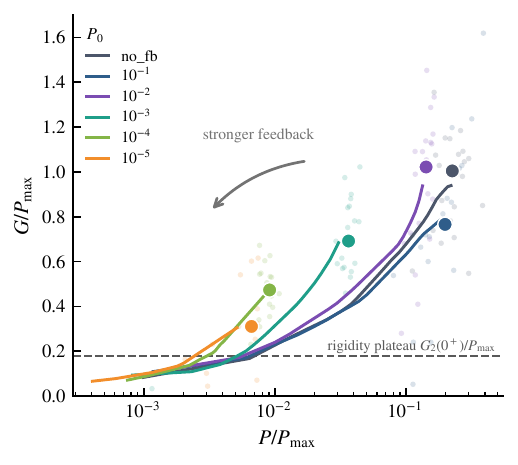}
\caption{
Post-jamming state diagram for $L=15D_0$. Each curve is the mean trajectory of the arrested packing in the $(P/P_{\max},\,G/P_{\max})$ plane as the packing fraction increases from $\phi_J$ to the depletion density $\phi_2$, at fixed feedback strength $P_0$. Filled markers denote the arrest endpoints $(P_2,G_2)$.
Rigidity plateau $G_2(0^+)=n_Jk\ell_c^2M_4u_J$ is denoted with a dashed line.
The plateau inputs are measured from the $P_0=10^{-5}$ jammed packings.}
\label{fig:statediagram}
\end{figure}

Relations~\eqref{eq:P2integral}--\eqref{eq:G2arrest} describe two nearly independent axes of the post-jamming state. Along the structural axis, the excess coordination $\Delta Z=\Delta Z_0+2(u_J-u)$ sets the shear rigidity and depends on feedback only weakly, through $\Delta Z_0(P_2)$, so the stiffness approaches the plateau $G_2(0^+)=n_Jk\ell_c^2M_4u_J$. By contrast, along the stress axis the arrest pressure is governed by the compressed-sector flux fraction $\lambda_c$ and is suppressed exponentially, $P_2\propto\exp(-P^\dagger/P_0)$. Figure~\ref{fig:statediagram} traces these post-jamming trajectories in the $(P,G)$ plane. 
As feedback strengthens, the arrest endpoints spread across more than two decades in pressure while approaching the rigidity plateau. Growth feedback therefore advances the arrest state almost vertically in this plane---adding rigidity---rather than horizontally along the pressure axis that controls ordinary compression-driven jamming.

Similar consequences of the contact excess can be observed in packings' spectral properties. Each unconstrained bud carries a floppy rotational mode and the arrested packing supports a band of quasi-zero-frequency vibrational modes whose count tracks $N_u=u\,N_{\nr}$, while the characteristic frequency above which excess modes appear is set by the modified excess coordination, $\omega^\ast\propto\Delta Z=\Delta Z_0+2(u_J-u)$, not the naive count. As the free-bud reservoir is depleted from $u_J$ to $\varepsilon$, this band drains and $\omega^\ast$ rises, as expected for marginal central-force solids~\cite{Wyart2005EPL,Silbert2005}. Appendix~\ref{app:dos} reports the vibrational density of states $D(\omega)$ ({\it cf.} Fig.~\ref{figS:dos}) obtained from the Hessian of the arrested configurations and tests the scaling of $\omega^\ast$ with the modified $\Delta Z$. This agreement between the corrected coordination excess $\Delta Z$ and characteristic frequency $\omega^\ast$ places the feedback-rigidified packings within the jamming universality class once the unconstrained-bud modes are properly identified~\cite{Brito2018,Ikeda2021}.

\runin{Conclusions} We have developed a theory of how growth feedback stiffens the packing after jamming, connecting proliferating active matter to jamming and rigidity in dense packings \cite{Weady2024,Hallatschek2023}. At jamming, a finite set of buds is still mechanically incomplete. Compressive stress slows growth in the compressed sector, so growth is biased toward the remaining underconstrained buds (see Appendix~\ref{app:growth}). As these buds become mechanically complete, they add more contacts than the modified isostatic count requires. The packing therefore moves away from marginality, even though little of the added volume loads the force network. This contact surplus stiffens the packing while keeping the growth-generated pressure low. 
With the numerically measured input of $S(a)$ and $\bar{\chi}(P_0)$, the model quantitatively predicts the depletion density and the arrest values of pressure and the elastic moduli without adjustable parameters.
Furthermore, feedback-regulated growth lets a proliferating solid set its rigidity and its internal stress almost independently. In principle, it can approach the rigidity plateau $G_2(0^+)$ while keeping the arrest pressure---and hence the elastic energy stored in the contact network---vanishingly small, so that the stiffness-to-stress ratio $G_2/P_2$ diverges as feedback strengthens. In this perspective, growth is a form of activity that can push a jammed packing away from marginality by using up residual underconstrained degrees of freedom, and the same mechanism should appear in other growing jammed systems, provided jamming leaves an underconstrained sector. This decoupling between rigidity and prestress is absent in compression-driven jamming and may provide a novel route to growth-driven rigidity in proliferating active matter.

\runin{Acknowledgments} We thank Emil \mbox{Albrychiewicz} for helpful comments on the manuscript.

\section*{Code Availability}
The simulation and analysis code needed to reproduce the results is available at \url{https://github.com/pgniewko/jamming-growth}.

\clearpage
\onecolumngrid
\appendix
\makeatletter
\@addtoreset{figure}{section}
\@addtoreset{table}{section}
\@addtoreset{equation}{section}
\makeatother
\renewcommand{\thefigure}{\Alph{section}\arabic{figure}}
\renewcommand{\thetable}{\Alph{section}\arabic{table}}
\renewcommand{\theequation}{\Alph{section}\arabic{equation}}

\section{Finite-size Maxwell counting and contact budget}\label{app:count}

\subsection{Modified Maxwell counting}
This work uses the total-count isostatic relation
\begin{equation}
N_c^{\iso}=6(N-N_f)-2N_u-2,
\label{eqS:codeiso}
\end{equation}
where $N_f$ is the number of rattlers, $N_u$ is the number of unconstrained buds among the remaining cells, and $N_c$ counts contacts twice. Let $N_{\nr}=N-N_f$ and $u=N_u/N_{\nr}$. The direct Maxwell count for a periodic system gives
\begin{equation}
N_c^{(0)}=2\bigl(3N_{\nr}-N_u-2\bigr)=6N_{\nr}-2N_u-4,
\label{eqS:Nc0}
\end{equation}
because each nonrattler budding cell has three rigid-body degrees of freedom in two dimensions, each unconstrained bud leaves one rotational mode effectively free, and the two global translations are removed. As in finite-size jamming of sphere packings \cite{Goodrich2012}, one additional physical contact is needed to generate the state of self-stress that supports compression. In doubled-contact counting this adds $+2$, which yields Eq.~\eqref{eqS:codeiso}. Dividing by $N_{\nr}$ gives
\begin{equation}
Z_{\iso}(u)=6-2u-\frac{2}{N_{\nr}},
\qquad
u\equiv \frac{N_u}{N_{\nr}}.
\label{eqS:Ziso}
\end{equation}

The structural relation used in the main text follows from the same contact budget. Let $\Delta Z\equiv Z-Z_{\iso}(u)$ be the excess coordination above the modified isostatic count. When one unconstrained bud becomes mechanically complete, the actual contact count increases by two physical contacts, or four counted contacts, whereas the modified isostatic count increases by only two counted contacts. In differential form,
\begin{equation}
\dd Z=-4\,\dd u,
\qquad
\dd Z_{\iso}=-2\,\dd u,
\label{eqS:dZdu}
\end{equation}
so
\begin{equation}
\dd(\Delta Z)=-2\,\dd u.
\label{eqS:dDZdu}
\end{equation}
Integrating from the jamming point, where the unconstrained-bud fraction is $u_J$, gives
\begin{equation}
\Delta Z(P,u)=\Delta Z_0(P)+2(u_J-u),
\label{eqS:DZtotal}
\end{equation}
with $\Delta Z_0(P)$ the no-feedback reference branch. The argument treats completion as the effective creation of two contacts, as in Ref.~\cite{Gniewek2019}. A bud can temporarily acquire one extra contact before it is fully constrained, but that intermediate state is not included in the present mean-field description.

\subsection{Geometric estimate of $u_J$}

The fraction of unconstrained buds at jamming $u_J$ is a central variable in the presented theory and is treated as a measured input. In this section, we present an estimate that turns this parameter from a measured one to a fixed one on the grounds of the mean-field theory. First, we note that by the counting argument presented above, the orientational coordinate of a cell is freed when its bud lobe carries no contact---the first bud contact removes the free rotation [Eq.~\eqref{eqS:Nc0}]. Within the mean-field count, a bud is therefore unconstrained if its lobe is contactless, so

\begin{equation}
u_J=\Pr(\text{bud lobe carries $0$ contacts at }\phi_J).
\label{eqS:ujdef}
\end{equation}

Placing bud contacts independently makes this zero-contact probability Poisson in the mean bud coordination $\bar z_b$,

\begin{equation}
u_J\simeq \ee^{-\bar z_b}.
\label{eqS:ujpoisson}
\end{equation}

At jamming the packing is marginal, $\Delta Z_0\to0$, so $Z_J=Z_{\iso}(u_J)=6-2u_J$. Because a cell comprises two lobes and $Z$ is the doubled contact count per cell, $Z=\bar z_m+\bar z_b$, with $\bar z_m$ and $\bar z_b$ the mean numbers of contacts carried by the mother and bud lobes. This yields the relation

\begin{equation}
\bar z_m+\bar z_b=6-2u_J.
\label{eqS:sumrule}
\end{equation}

A mother lobe is a fully embedded $2$D disk, whose geometric contact number we approximate with the disk isostatic value $\bar z_m\approx2d=4$. Then $\bar z_b=2-2u_J$, and Eq.~\eqref{eqS:ujpoisson} gives:

\begin{equation}
u_J=\ee^{-(2-2u_J)}
\;\Longleftrightarrow\;
u_J\,\ee^{-2u_J}=\ee^{-2}
\;\Longrightarrow\;
u_J=-\tfrac12\,W_0\!\bigl(-2\ee^{-2}\bigr)\approx0.20,
\label{eqS:ujsolve}
\end{equation}

with $W_0$ the principal branch of the Lambert function. This parameter-free estimate is close to the measured value $u_J\simeq0.23$ of the $L=15D_0$ packings.

\section{\texorpdfstring{Finite-size effects}{Finite-size effects}}\label{app:sizeeffects}

\begin{figure*}[t]
\centering
\smartincludegraphics[width=0.92\textwidth]{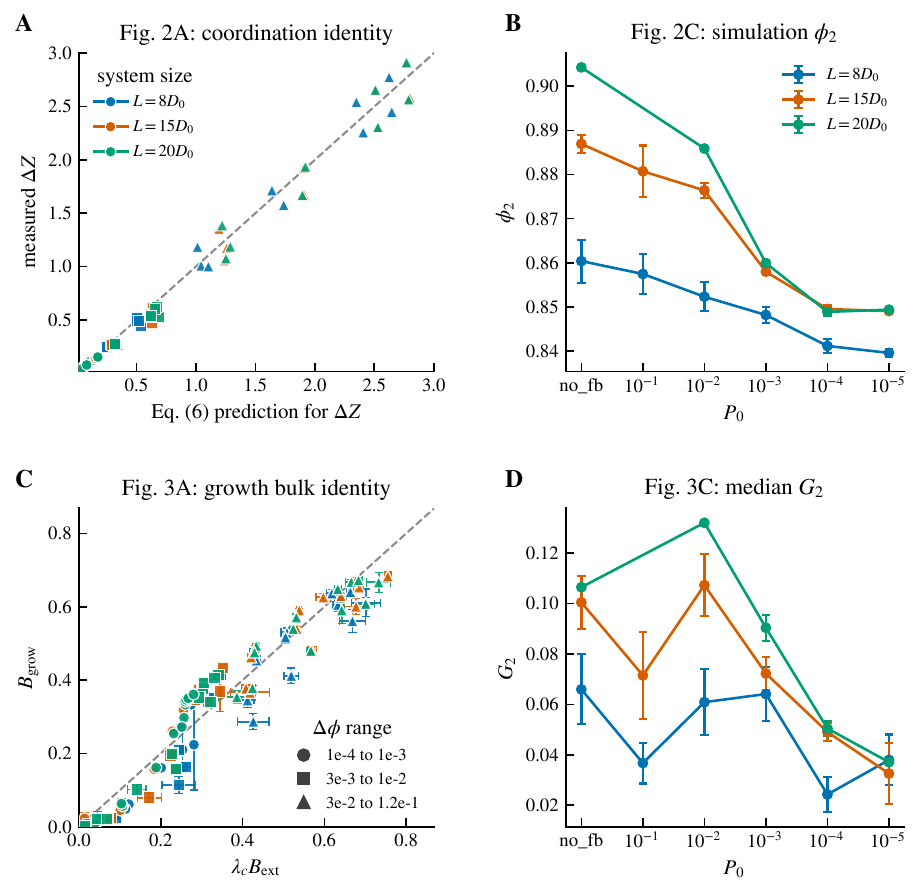}
\caption{
Finite-size comparison for $L=8D_0$, $15D_0$, and $20D_0$.
(\textbf{A}) Measured excess coordination versus the counting prediction from Fig.~\ref{fig:structure}(\textbf{A}).
(\textbf{B}) Simulation-measured depletion density $\phi_2(P_0)$ from Fig.~\ref{fig:structure}(\textbf{C}), without the theory branch.
(\textbf{C}) Measured growth bulk response versus $\lambda_cB_{\ext}$ from Fig.~\ref{fig:mechanics}(\textbf{A}).
(\textbf{D}) Median shear modulus at the depletion point from Fig.~\ref{fig:mechanics}(\textbf{C}).
Error bars have the same meaning as in the main text figures.
}
\label{figS:sizeeffects}
\end{figure*}

The main figures are generated from the $L=15D_0$ dataset. To test that the conclusions are not tied to the chosen system size $L$, Fig.~\ref{figS:sizeeffects} repeats the analysis behind Fig.~\ref{fig:structure}(\textbf{A}), the simulation branch of Fig.~\ref{fig:structure}(\textbf{C}), Fig.~\ref{fig:mechanics}(\textbf{A}), and the median line in Fig.~\ref{fig:mechanics}(\textbf{C}) for $L=8D_0$, $15D_0$, and $20D_0$. In these comparisons the point color denotes system size rather than feedback strength $P_0$.

The structural identity $\Delta Z=\Delta Z_0+2(u_J-u)$ and the growth-response identity $B_{\grow}=\lambda_cB_{\ext}$ agree across the three sizes within the expected range. 
The smallest systems show the largest variation, as expected from the $N_{\nr}^{-1}$ term in the modified isostatic count. 
The measured depletion density $\phi_2$ and the shear modulus $G_2$ at $\phi_2$ also show no significant drift between $L=15D_0$ and $20D_0$ over the feedback range. Notably, they approach the same limiting values as $P_0\rightarrow0^+$.

\section{Depletion beyond the sharp-threshold limit}\label{app:depletion}

{
\subsection{Threshold broadening}

Depletion of the buds that are unconstrained at jamming is modeled by introducing the completion-threshold survival function $S(a)$, where $a$ is the added bud area measured relative to the state at $\phi_J$. In the no-injection limit, the surviving unconstrained-bud fraction is
\begin{equation}
u(a)=u_J S(a).
\label{eqS:uofa}
\end{equation}
If the compressed-sector mean growth factor varies slowly over the depletion interval, the density increment obeys
\begin{equation}
\frac{\dd\phi}{\dd a}
=
n_J\left[u(a)+(1-u(a))\bar\chi(P_0)\right],
\label{eqS:dphida}
\end{equation}
where $n_J$ is the nonrattler density at jamming and $\bar\chi(P_0)$ is the depletion-interval average of the compressed-sector growth factor. Integrating Eq.~\eqref{eqS:dphida} gives
\begin{equation}
\phi(a;P_0)-\phi_J
=
n_J\left[\bar\chi(P_0)a+\bigl(1-\bar\chi(P_0)\bigr)u_J\int_0^a S(a')\,\dd a'\right].
\label{eqS:phiofa}
\end{equation}

The main text defines $\phi_2$ operationally through $u(\phi_2,P_0)=\varepsilon$, with $\varepsilon$ a small cutoff. For the finite systems analyzed here, $\varepsilon$ is naturally of order $N_{\nr,J}^{-1}$, so this cutoff corresponds to exhaustion of the last initially unconstrained bud. To quantify how a finite threshold width broadens that point, write
\begin{equation}
a=a_c+\sigma_a x,
\qquad
S(a)=\mathcal S(x),
\label{eqS:scaledS}
\end{equation}
where $a_c$ is the mean completion area and $\sigma_a$ is the edge width. Substituting into Eq.~\eqref{eqS:phiofa} and expanding around the solution $x_2$ of $u_J\mathcal S(x_2)=\varepsilon$ yields
\begin{equation}
\phi-\phi_2
\simeq
n_J\bigl[\varepsilon+(1-\varepsilon)\bar\chi(P_0)\bigr]\sigma_a(x-x_2),
\label{eqS:phiwidth}
\end{equation}
so the density broadening is
\begin{equation}
\delta\phi_2
\sim
n_J\bigl[\varepsilon+(1-\varepsilon)\bar\chi(P_0)\bigr]\sigma_a.
\label{eqS:widthphi2}
\end{equation}
Therefore, a finite edge width in the threshold distribution $S(a)$ leads directly to a finite width in the approach to $\phi_2$. If instead the survival function decays to zero like a power law,
\begin{equation}
S(a)\sim C(a_c-a)^\beta,
\qquad a\to a_c^{-},
\label{eqS:edgebeta}
\end{equation}
then near $a_c$ one has $u(a)=u_JC(a_c-a)^\beta$. For the finite-cutoff definition used here, $u(\phi_2,P_0)=\varepsilon$ gives
\begin{equation}
\phi_2-\phi
\simeq
n_J\bigl[\varepsilon+(1-\varepsilon)\bar\chi(P_0)\bigr](a_2-a),
\label{eqS:phi2minusphi}
\end{equation}
with $a_2$ fixed by $u(a_2)=\varepsilon$. Hence
\begin{equation}
u(\phi)-\varepsilon \propto \phi_2-\phi,
\qquad
\Delta Z_2-\Delta Z(\phi)\propto \phi_2-\phi,
\label{eqS:continuousedge}
\end{equation}
when $\bar\chi>0$. By contrast, the formal exact-depletion limit $\varepsilon\to 0$ recovers
\begin{equation}
u(\phi)\propto (\phi_{\mathrm{dep}}-\phi)^\beta,
\qquad
\Delta Z_{\mathrm{dep}}-\Delta Z(\phi)\propto (\phi_{\mathrm{dep}}-\phi)^\beta,
\label{eqS:exactdep}
\end{equation}
where $\phi_{\mathrm{dep}}$ is the corresponding exact-depletion density. The case $\bar\chi=0$ gives a different scaling and is not discussed here.

\subsection{Injection of new degrees of freedom by continued division}

The main text analysis focuses on the {\it no-injection} regime, in which free buds are only depleted by mechanical completion. Continued division can be included by adding a source term for newly created unconstrained buds,
\begin{equation}
\frac{\dd u}{\dd a}
=
-h(a)u+I_{\divv}(a,P_0),
\qquad
h(a)\equiv -\frac{\dd}{\dd a}\ln S(a),
\label{eqS:sourceSink}
\end{equation}
where $h(a)$ is the completion hazard associated with the no-injection survival function $S(a)$ and $I_{\divv}$ is the injection rate.

A simple solvable limit is
\begin{equation}
h(a)=h_0,
\qquad
I_{\divv}(a,P_0)=I_0(P_0),
\label{eqS:constinj}
\end{equation}
for which $S(a)=\ee^{-h_0a}$. With the initial condition $u(0)=u_J$, Eq.~\eqref{eqS:sourceSink} gives
\begin{equation}
u(a)=u_J\ee^{-h_0a}+\frac{I_0(P_0)}{h_0}\bigl(1-\ee^{-h_0a}\bigr).
\label{eqS:uinjsolution}
\end{equation}
At large $a$, the unconstrained-bud fraction saturates at
\begin{equation}
u_{a\gg 0}=\frac{I_0(P_0)}{h_0}.
\label{eqS:injplateau}
\end{equation}
If $I_0/h_0\ll \varepsilon$, the cutoff $u=\varepsilon$ is still reached and the no-injection theory remains accurate up to small corrections. If $I_0/h_0$ is comparable to $\varepsilon$, $u(a)$ saturates before exact depletion is reached, and $\phi_2$ is replaced by a rounded crossover.

We measured this correction directly by counting all post-jamming division events in the saved lineage data. For each run, the cumulative count
$N_{\divv}^{\mathrm{all}}(\Delta\phi)$ includes every division event after $\phi_J$, with $\Delta\phi=\phi-\phi_J$, and $\phi_2$ is identified as the first saved state at which the population of buds that were unconstrained at jamming is exhausted. Fig.~\ref{figS:lineagedivisions} shows the result. The number of new divisions before $\phi_2$ is substantial for no feedback and weak feedback, with median values $N_{\divv}^{\mathrm{all}}(\phi_2)/N_J\simeq0.081$, $0.071$, and $0.053$ for no feedback, $P_0=10^{-1}$, and $P_0=10^{-2}$, respectively. It drops to $0.013$ at $P_0=10^{-3}$ and to approximately $3\times10^{-3}$ or less for $P_0=10^{-4}$--$10^{-5}$. Thus the no-injection approximation is quantitatively most accurate in the strong-feedback regime that controls the low-$P_0$ rigidity plateau.

\begin{figure}[t]
\centering
\smartincludegraphics[width=0.92\columnwidth]{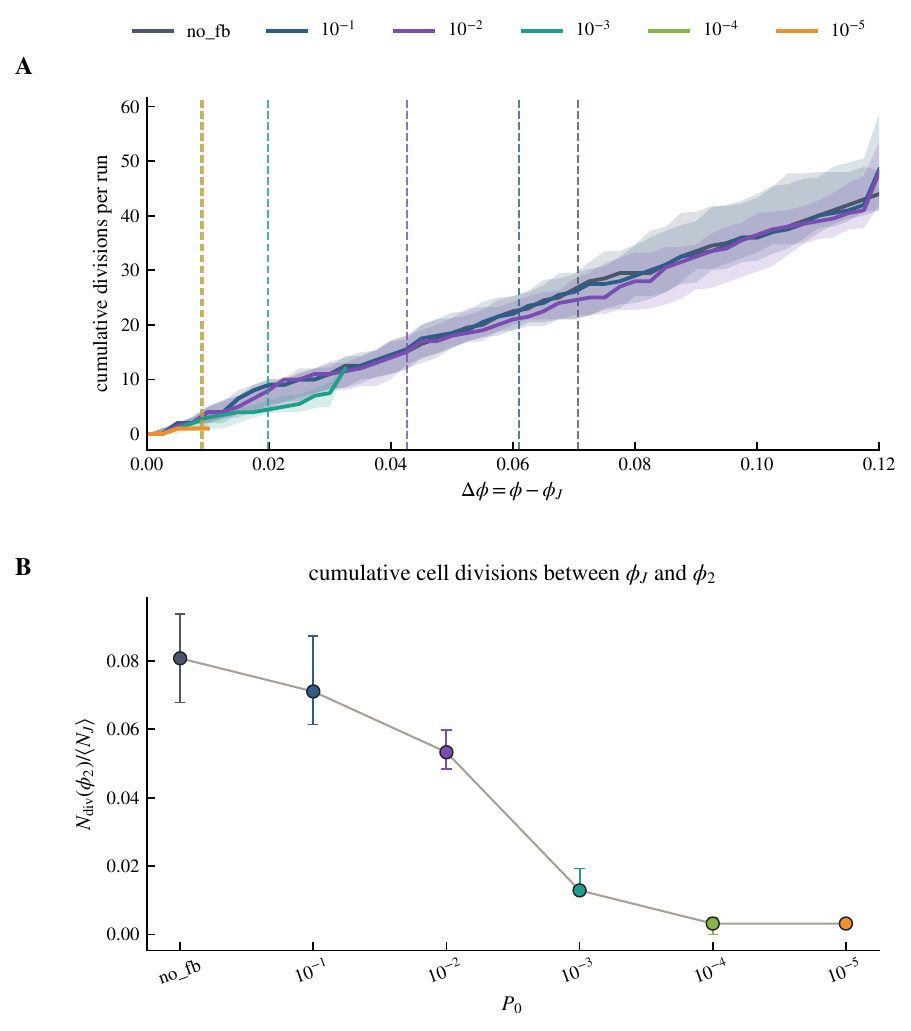}
\caption{
Post-jamming division counts.
(\textbf{A}) Median cumulative number of all division events after $\phi_J$, with interquartile bands across runs. Dashed vertical lines mark the median observed $\Delta\phi_2=\phi_2-\phi_J$ for each feedback strength.
(\textbf{B}) Median number of divisions accumulated by $\phi_2$, normalized by the mean number of cells at jamming. Division-mediated injection is appreciable for no feedback and weak feedback but is strongly suppressed for $P_0\le10^{-3}$. The error bars are asymmetric 95\% bootstrap confidence intervals for the median.
}
\label{figS:lineagedivisions}
\end{figure}
}

\section{\texorpdfstring{Affine and excess-mode mechanics}{Affine and excess-mode mechanics}}\label{app:mechanics}

{
\subsection{Bulk modulus}

We first derive the instantaneous bulk response of a mechanically arrested packing, assuming affine deformations. Consider a small isotropic areal strain $\eta\equiv \delta\phi/\phi$, so that the strain tensor in two dimensions is $\bm\epsilon=-(\eta/2)\mathbf I$. For a contact of pre-strain length $\ell$ and unit normal $\bm n$, the affine normal strain is
\begin{equation}
\frac{\delta\ell}{\ell}
=
\bm n\cdot\bm\epsilon\cdot\bm n
=
-\frac{\eta}{2}.
\label{eqS:deltaellbulk}
\end{equation}
The elastic energy stored in one contact is therefore
\begin{equation}
E_{\mathrm{cont}}^{\mathrm{aff}}
=
\frac{k}{2}(\delta\ell)^2
=
\frac{k\ell^2}{8}\eta^2.
\label{eqS:Econtbulk}
\end{equation}
With contact density $\rho_c=n_JZ/2$, and replacing $\ell^2$ by its contact-averaged value $\ell_c^2$, the affine energy density becomes
\begin{equation}
\delta\mathcal E_{\mathrm{bulk}}
=
\rho_c\avg{E_{\mathrm{cont}}^{\mathrm{aff}}}
=
\frac{n_Jk\ell_c^2}{16}Z\,\eta^2.
\label{eqS:Ebulkdens}
\end{equation}
Identifying $\delta\mathcal E_{\mathrm{bulk}}=\tfrac12 B_{\ext}\eta^2$ gives
\begin{equation}
B_{\ext}=\frac{n_Jk\ell_c^2}{8}\,Z.
\label{eqS:Bext}
\end{equation}

\noindent\textit{Affine character of the bulk estimate.}---Equation~\eqref{eqS:Bext} keeps the contact network fixed and lets each contact follow the imposed isotropic strain. A fully relaxed packing also has nonaffine particle motions, so the measured bulk modulus is slightly softer than this affine value. In jammed packings, however, compression remains nearly affine: isotropic loading makes all contacts carry the same sign of strain, leaving little room for the nonaffine relaxations that soften the shear response. As a result, the nonaffine correction mainly changes the prefactor and leaves the leading scaling, $B\simeq B_{\ext}\propto Z$, unchanged~\cite{Ellenbroek2009}.

\subsection{Shear modulus}

For simple shear, write the contact normal as $\bm n=(\cos\theta,\sin\theta)$. The orientational fourth moment relevant to shear is
\begin{equation}
M_4\equiv \avg{n_x^2n_y^2}
=
\frac{1}{2\pi}\int_0^{2\pi}\cos^2\theta\sin^2\theta\,\dd\theta
=
\frac18.
\label{eqS:M4}
\end{equation}

Consider a central-force contact of length $\ell$ under a small simple shear strain $\gamma$, with displacement field $\bm u=(\gamma y,0)$. The symmetric strain tensor is
\begin{equation}
\bm\epsilon
=
\frac12\bigl(\nabla\bm u+\nabla\bm u^{\mathsf T}\bigr)
=
\frac{\gamma}{2}
\begin{pmatrix}
0 & 1\\
1 & 0
\end{pmatrix}.
\label{eqS:epsshear}
\end{equation}
The affine normal extension of the contact is then
\begin{equation}
\frac{\delta\ell}{\ell}
=
\bm n\cdot\bm\epsilon\cdot\bm n
=
\gamma n_x n_y,
\label{eqS:deltaellshear}
\end{equation}
so the affine elastic energy stored in one contact is
\begin{equation}
E_{\mathrm{cont}}^{\mathrm{aff}}
=
\frac{k}{2}(\delta\ell)^2
=
\frac{k\ell^2}{2}\gamma^2 n_x^2n_y^2.
\label{eqS:Econtshear}
\end{equation}

Near marginality, nonaffine relaxation removes the contribution of isostatic contacts to the shear modulus. The remaining energy is carried by the excess states of self-stress, whose number scales with $\Delta Z$ in central-force packings \cite{Wyart2005EPL,Ellenbroek2006,Goodrich2012}. 
We estimate that response by assigning each excess contact the affine energy of Eq.~\eqref{eqS:Econtshear} with the isotropic angular weight $M_4$---an effective-medium approximation in which the excess sector is treated as an affine central-force network. Using the excess contact density
\begin{equation}
\rho_c^{\mathrm{exc}}=\frac{n_J}{2}\Delta Z,
\label{eqS:rhocexc}
\end{equation}
the shear energy density becomes
\begin{equation}
\delta\mathcal E_{\shear}
=
\frac{k\ell_c^2}{2}\gamma^2 M_4\rho_c^{\mathrm{exc}}
=
\frac{n_Jk\ell_c^2}{4}M_4\Delta Z\,\gamma^2.
\label{eqS:Eshear}
\end{equation}
Identifying $\delta\mathcal E_{\shear}=\tfrac12 G\gamma^2$ gives
\begin{equation}
G=\frac{n_Jk\ell_c^2}{2}M_4\Delta Z.
\label{eqS:G}
\end{equation}
Equation~\eqref{eqS:G} gives a leading-order estimate near marginality, with residual contact anisotropy and higher-order nonaffine corrections not treated here. For an isotropic network with $M_4=1/8$, Eq.~\eqref{eqS:G} agrees quantitatively with the Zaccone--Scossa-Romano expression for two-dimensional packings with repulsive central forces in Refs.~\cite{Zaccone2011,Zaccone2025JAP}.

\subsection{Flux partition, arrest pressure, and plateau moduli}

During a small growth increment, let each unconstrained bud gain added area $\dd a$. Approximating the nonrattler density by $n_J$, the unconstrained sector contributes $n_Ju\,\dd a$ to $\dd\phi$, while the compressed sector contributes $n_J(1-u)\chi_c\,\dd a$. Summing the two gives
\begin{equation}
\dd\phi
=
n_J\bigl[u+(1-u)\chi_c\bigr]\dd a.
\label{eqS:dphi}
\end{equation}
The fraction of the total density increment that enters the compressed sector is therefore
\begin{equation}
\lambda_c
=
\frac{(1-u)\chi_c}{u+(1-u)\chi_c}.
\label{eqS:lambdac}
\end{equation}

Because unconstrained buds grow into available void space without loading existing contacts, only the compressed-sector increment directly loads the force-bearing network. Writing $\dd\phi_c=\lambda_c\,\dd\phi$ and using the bulk modulus from Eq.~\eqref{eqS:Bext},
\begin{equation}
\dd P
=
\frac{B_{\ext}}{\phi}\dd\phi_c
=
\frac{B_{\ext}}{\phi}\lambda_c\,\dd\phi.
\label{eqS:dP}
\end{equation}
\begin{equation}
B_{\grow}\equiv \phi\left(\frac{\partial P}{\partial\phi}\right)_{\grow}
=
\lambda_c B_{\ext}.
\label{eqS:Bgrow}
\end{equation}
Accordingly, Eq.~\eqref{eqS:Bgrow} is the leading-order sector-partition prediction for the elastic growth response. Integrating from $\phi_J$ to $\phi_2$ gives the arrest pressure
\begin{equation}
P_2(P_0)
=
\int_{\phi_J}^{\phi_2(P_0)}
\frac{\dd\phi}{\phi}\,\lambda_c(\phi,P_0)B_{\ext}(\phi,P_0).
\label{eqS:P2}
\end{equation}

In the sharp-threshold limit, the compressed sector receives the density increment
\begin{equation}
\delta\phi_c\simeq n_Ja_c(1-u_J)\bar\chi(P_0).
\label{eqS:dphic}
\end{equation}
If $B_{\ext}$ varies slowly over that interval,
\begin{equation}
P_2(P_0)
\simeq
\frac{B_2(0^+)}{\phi_J}\,n_Ja_c(1-u_J)\bar\chi(P_0).
\label{eqS:P2strong}
\end{equation}
If the compressed-sector pressure is concentrated near a typical value $P^\dagger$, then
\begin{equation}
\bar\chi(P_0)\simeq \exp\!\left[-\frac{P^\dagger}{P_0}\right],
\qquad
P_2(P_0)\propto \exp\!\left[-\frac{P^\dagger}{P_0}\right].
\label{eqS:P2scaling}
\end{equation}

At the depletion point $u(\phi_2,P_0)=\varepsilon$, the coordination number and excess coordination are
\begin{equation}
Z_2=6+2u_J-4\varepsilon+\Delta Z_0(P_2),
\qquad
\Delta Z_2=2(u_J-\varepsilon)+\Delta Z_0(P_2),
\label{eqS:Z2DZ2}
\end{equation}
which give
\begin{equation}
B_2(P_0)
=
\frac{n_Jk\ell_c^2}{8}\Bigl[6+2u_J-4\varepsilon+\Delta Z_0(P_2)\Bigr],
\label{eqS:B2}
\end{equation}
\begin{equation}
G_2(P_0)
=
\frac{n_Jk\ell_c^2}{2}M_4\Bigl[2(u_J-\varepsilon)+\Delta Z_0(P_2)\Bigr].
\label{eqS:G2}
\end{equation}
In the low-pressure strong-feedback limit, these reduce to
\begin{equation}
B_2(0^+)\simeq \frac{n_Jk\ell_c^2}{8}\bigl(6+2u_J\bigr),
\label{eqS:B2plateau}
\end{equation}
\begin{equation}
G_2(0^+)\simeq n_Jk\ell_c^2M_4u_J,
\label{eqS:G2plateau}
\end{equation}
up to the small cutoff and no-feedback compression corrections.

In the strong-feedback limit, $G_2$ approaches the finite plateau of Eq.~\eqref{eqS:G2plateau} while $P_2$ vanishes exponentially as in Eq.~\eqref{eqS:P2scaling}. The ratio $G_2/P_2$ therefore diverges as $P_0\to 0$. Fig.~\ref{figS:mechanicspaired}(\textbf{B}) shows this divergence in the simulation data.

The confluence pressure scale used in the main text can be estimated by affinely compressing a monodisperse packing from $\phi_J\simeq0.84$ to $\phi=1$. This gives
\begin{equation}
\frac{P_{\max}}{k}\simeq \frac{2}{\pi}(1-\phi_J)\approx 0.1.
\label{eqS:Pmax}
\end{equation}

\begin{figure}[t]
\centering
\smartincludegraphics[width=0.90\columnwidth]{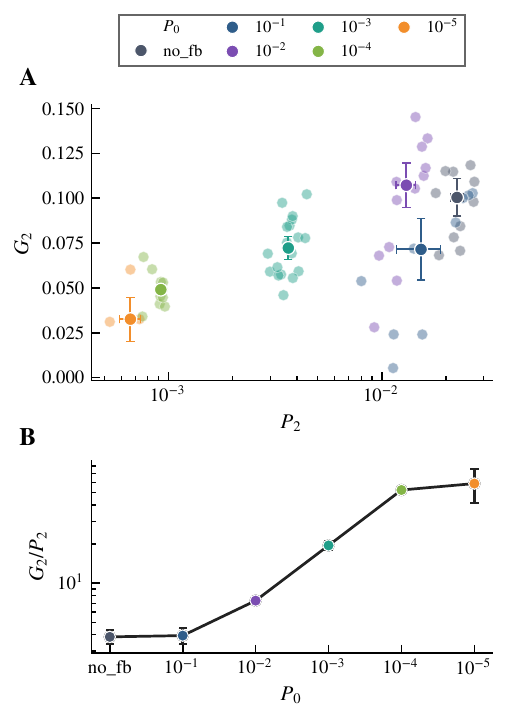}
\caption{
Additional arrest-state statistics at the depletion point $\phi_2$.
(\textbf{A}) Arrest states in the $(P_2,G_2)$ plane. Large markers denote the corresponding per-$P_0$ medians with bootstrap median errors.
(\textbf{B}) Median of the ratio $G_2/P_2$ for the same arrest states. Error bars are SE of the median.
}
\label{figS:mechanicspaired}
\end{figure}
}

\section{Growth-rate crossover}\label{app:growth}

{
\subsection{Distribution and heterogeneity}

The normalized growth rate $g\equiv \gamma/\gamma_0$ has a two-component distribution. If $\pi_c(p)$ is the compressed-sector pressure distribution, then for $g=\exp(-p/P_0)$, the compressed-sector distribution follows from the change of variables $p=-P_0\ln g$ and $|\dd p/\dd g|=P_0/g$, i.e.
\begin{equation}
\Pi_c(g)=\frac{P_0}{g}\pi_c(-P_0\ln g),
\qquad 0<g\le 1.
\end{equation}
Unconstrained buds contribute a peak at $g=1$, so $\Pi_u(g)=\delta(g-1)$. Adding these two terms, for an unconstrained-bud fraction $u$, gives:
\begin{equation}
\Pi(g\mid \phi,P_0)
=
u\,\delta(g-1)
+
(1-u)\frac{P_0}{g}\pi_c\!\left(-P_0\ln g\right),
\qquad
0<g\le 1.
\label{eqS:hist}
\end{equation}

The compressed-sector contribution is centered near $g_\ast=\exp(-P_\ast/P_0)$ if the pressure distribution is concentrated near a characteristic value $P_\ast$. The distribution is therefore bimodal when
\begin{equation}
P_0\ll P_\ast,
\label{eqS:bimodalcriterion}
\end{equation}
and becomes unimodal once $P_0\sim P_\ast$, when the compressed sector overlaps the unconstrained-bud peak near $g=1$.

A convenient scalar summary is the normalized growth-rate heterogeneity
\begin{equation}
h_\gamma\equiv \frac{\Var(g)}{\avg{g}^2}.
\label{eqS:hgamma}
\end{equation}
The moments of the full distribution are
\begin{equation}
\avg{g}=u+(1-u)\chi_c,
\qquad
\chi_c\equiv \avg{\ee^{-p/P_0}}_c,
\label{eqS:gmean}
\end{equation}
\begin{equation}
\avg{g^2}=u+(1-u)\chi_{c,2},
\qquad
\chi_{c,2}\equiv \avg{\ee^{-2p/P_0}}_c,
\label{eqS:g2mean}
\end{equation}
so
\begin{equation}
h_\gamma=
\frac{u+(1-u)\chi_{c,2}-\bigl[u+(1-u)\chi_c\bigr]^2}
{\bigl[u+(1-u)\chi_c\bigr]^2}.
\label{eqS:hgammafinal}
\end{equation}

\subsection{Numerical evidence}

At fixed $\Delta\phi=3\times10^{-3}$, we measure both the normalized heterogeneity $h_\gamma$ and the full histogram $\Pi(g)$. Equation~\eqref{eqS:hist} predicts that decreasing $P_0$ should drive the compressed-sector contribution toward smaller $g$ while leaving the unconstrained-bud peak pinned at $g=1$.

\begin{figure}[t]
\centering
\smartincludegraphics[width=0.97\columnwidth]{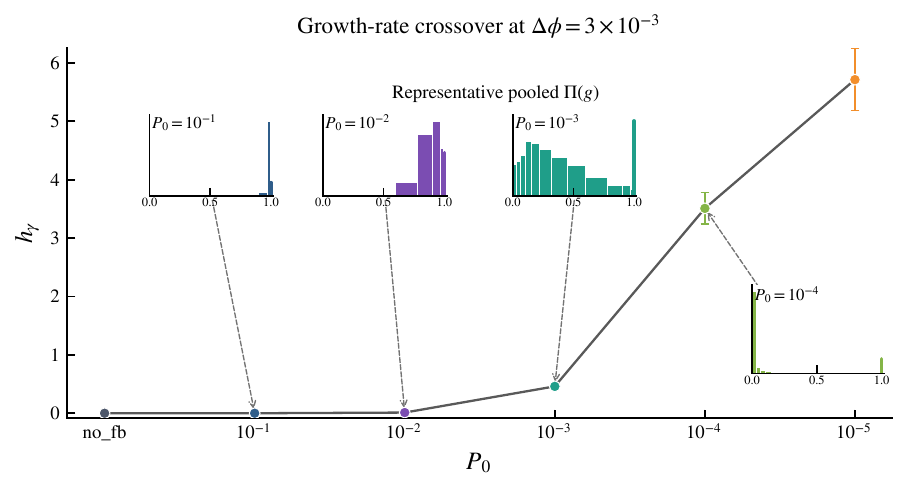}
\caption{
Growth-rate crossover for $L=15D_0$ at fixed $\Delta\phi=3\times10^{-3}$. The main curve shows the normalized heterogeneity $h_\gamma$ as a function of feedback strength $P_0$. Insets show representative histograms for $P_0=10^{-1},10^{-2},10^{-3},10^{-4}$, with arrows linking each histogram to its location on the main curve. The no-feedback reference is included only on the main curve, where the distribution collapses to $\delta(g-1)$. Error bars are bootstrap SE of the median across runs. 
}
\label{figS:growthcrossover}
\end{figure}

Numerical results presented in Fig.~\ref{figS:growthcrossover} agree with this prediction. For weak feedback ($P_0=10^{-1}$), the heterogeneity is essentially zero ($h_\gamma=1.4\times10^{-4}$), while the no-feedback reference on the main curve gives $h_{\gamma}=0$. In both cases the distribution is pinned at $g\approx1$. Heterogeneity then rises rapidly as feedback strengthens, with the crossover occurring around $P_0\sim10^{-2}$--$10^{-3}$. Notably, this is also the range in which the contribution from the compressed sector separates clearly from the unconstrained-bud peak at $g=1$.

In parallel, the increase in $h_\gamma$ is driven mainly by the displacement of the compressed-sector contribution rather than by a dramatic loss of unconstrained buds. The weight of the peak at $g=1$ is about $0.21$ at $P_0=10^{-1}$ and $P_0=10^{-2}$, and it is still about $0.16$ at $P_0=10^{-4}$. By contrast, the compressed-sector peak shifts from $g\simeq0.97$ at $P_0=10^{-1}$ to $g\simeq0.90$ at $P_0=10^{-2}$, then to $g\simeq0.17$ at $P_0=10^{-3}$, and finally to the lowest-$g$ bin by $10^{-4}$.

\subsection{Order-of-magnitude estimate for the crossover scale}

The crossover scale $P^\ast$ in Eq.~\eqref{eqS:bimodalcriterion} can be estimated from the elastic response to cellular growth. Strong suppression of compressed-sector growth begins once the growth rate drops by an order of magnitude
\begin{equation}
\exp\!\left[-\frac{p_{\bud}}{P_0}\right]\sim 0.1,
\label{eqS:suppression}
\end{equation}
which implies
\begin{equation}
P^\ast \sim \frac{p_{\bud}^{\mathrm{typ}}}{\ln 10}.
\label{eqS:pstarcriterion}
\end{equation}
A characteristic compressed-bud pressure can be estimated as
\begin{equation}
p_{\bud}^{\mathrm{typ}}\sim B_2(0^+)\,\varepsilon_c,
\label{eqS:pbudtyp}
\end{equation}
where $\varepsilon_c$ is the characteristic local strain needed to complete a compressed bud. If that strain is estimated from the characteristic added completion area $a_c$,
\begin{equation}
\varepsilon_c\sim \frac{a_c}{\ell_c^2},
\label{eqS:epsc}
\end{equation}
then, with $\ell_c$ the same contact-length scale that enters the elastic moduli, Eq.~\eqref{eqS:B2plateau} gives
\begin{equation}
\frac{P^\ast}{k}
\sim
\frac{1}{\ln 10}\frac{B_2(0^+)}{k}\frac{a_c}{\ell_c^2}
\sim
\frac{n_J(6+2u_J)}{8\ln 10}\,a_c.
\label{eqS:pstarbulkestimate}
\end{equation}

Using strong-feedback estimates from the $L=15D_0$ dataset, $n_J\simeq0.775$ and $u_J\simeq0.23$, together with completion thresholds in the range $a_c\sim4.7\times10^{-3}$ to $1.6\times10^{-2}$, Eq.~\eqref{eqS:pstarbulkestimate} gives
\begin{equation}
\frac{P^\ast}{k}\sim 1.3\times10^{-3}\ \text{to}\ 4.3\times10^{-3}.
\label{eqS:pstarrange}
\end{equation}
This matches the order of magnitude at which Fig.~\ref{figS:growthcrossover} shows the onset of clear bimodality. 
}

\section{\texorpdfstring{Vibrational density of states}{Vibrational density of states}}\label{app:dos}

\subsection{Hessian and soft-mode count}\label{app:hess}

We diagonalize the vibrational spectrum of the mechanically stable configurations to test the structural picture spectrally, following the vibrational characterization of the unjamming transition~\cite{Silbert2005}. The mother and bud lobes are rigidly fixed along the cell axis, so a bud carries no independent hinge coordinate. We discard rattlers and assign each nonrattler cell the three rigid-body coordinates $q=(x,y,\theta)$, where $(x,y)$ is the center-of-mass (COM) of the mother--bud dumbbell and $\theta$ is its orientation. The elastic energy is the sum over inter-cell lobe--lobe contacts of harmonic springs,
\begin{equation}
U=\sum_{\langle ij\rangle}\frac{k}{2}\,\delta_{ij}^2,
\qquad
\delta_{ij}=\bigl(R_i+R_j-r_{ij}\bigr)\Theta\!\bigl(R_i+R_j-r_{ij}\bigr),
\label{eqS:Uharm}
\end{equation}
with $R_i$ the lobe radii and $r_{ij}$ the lobe-center separation. The dynamical matrix is the mass-weighted Hessian
\begin{equation}
\mathcal D_{ab}=\frac{1}{\sqrt{m_a m_b}}\frac{\partial^2 U}{\partial q_a\,\partial q_b}\Bigg|_{\rm arrest},
\label{eqS:dynmat}
\end{equation}
evaluated at the force-balanced state and including both the spring-stiffness (acting along the inter-lobe contact line $\hat{n}\hat{n}^{\top}$) and the prestress (initial-stress acting transverse to the contact line $1-\hat{n}\hat{n}^{\top}$) contributions, since the latter is not negligible near marginality. Lobe masses are assigned at fixed areal density---we give the mother lobe unit mass and the bud lobe mass $m_{\rm bud}=(R_{\rm bud}/R_{\rm mother})^2$. The translational coordinates of a cell thus carry mass $1+(R_{\rm bud}/R_{\rm mother})^2$ and the orientation $\theta$ the rigid-body moment of inertia $\mathcal I=\sum_a m_a\!\left(\ell_a^2+\tfrac12 R_a^2\right)$ (parallel-axis plus disk self-spin, with $\ell_a$ the offset of lobe $a$ from the dumbbell COM). We solve the generalized problem $\mathcal H\,v=\omega^2\mathcal M\,v$ with $\mathcal H$ the full Hessian (built by finite differences of the analytic gradient on the fixed contact network) and $\mathcal M$ the consistent dumbbell mass matrix, and remove the two uniform-translation zero modes. Its eigenvalues $\lambda_n$ give the mode frequencies $\omega_n=\sqrt{\lambda_n}$, and the density of states is
\begin{equation}
D(\omega)=\frac{1}{N_{\rm dof}}\sum_n \delta(\omega-\omega_n).
\label{eqS:dos}
\end{equation}

Each unconstrained bud leaves the cell-orientation coordinate $\theta$ of its dumbbell nearly free---when no contacts load the bud the mother lobe sits close to the dumbbell COM, so rotating $\theta$ pivots the contact-free bud at almost no energy cost. The number of quasi-zero modes (eigenvalues below a small cutoff $\lambda<\lambda_{\rm tol}$) is then predicted to track the unconstrained-bud count,
\begin{equation}
N_{\rm zero}\simeq N_u = u\,N_{\nr}.
\label{eqS:nzero}
\end{equation}
The characteristic frequency above which excess modes proliferate is set by the modified excess coordination, $\Delta Z$, measured against the \emph{modified} isostatic count $Z_{\iso}(u)=6-2u$ rather than against a fixed $Z_{\rm iso}^{\rm naive}=2f=2\times3=6$ (three DOF per dumbbell):
\begin{equation}
\omega^\ast\propto \Delta Z=\Delta Z_0(P)+2(u_J-u),
\label{eqS:omegastar}
\end{equation}
Equation \eqref{eqS:omegastar} can be interpreted as the spectral counterpart of the structural identity Eq.~\eqref{eqS:DZtotal}---as the free-bud reservoir is depleted from $u_J$ to the cutoff $\varepsilon$, the quasi-zero band of Eq.~\eqref{eqS:nzero} drains and $\omega^\ast$ rises.

\subsection{Vibrational density of states\texorpdfstring{---}{---}numerical results}

\begin{figure}[t]
\centering
\smartincludegraphics[width=0.97\columnwidth]{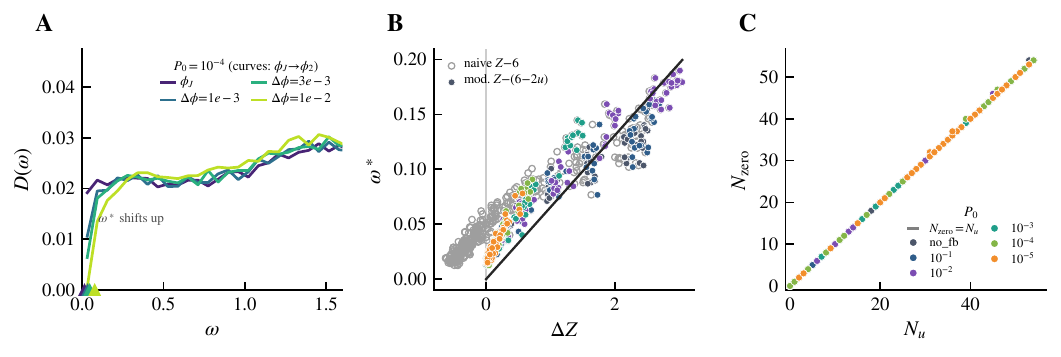}
\caption{
Vibrational density of states for $L=15D_0$.
(\textbf{A}) $D(\omega)$ at several packing fractions from $\phi_J$ to $\phi_2$ at fixed strong feedback. We observe that the low-frequency spectral weight drains and the band edge $\omega^\ast$ (triangles) shifts upward as unconstrained buds are completed (see Appendix \ref{app:hess}).
(\textbf{B}) Characteristic frequency $\omega^\ast$ versus the modified excess coordination $\Delta Z=\Delta Z_0+2(u_J-u)$, tested over $(P_0,\Delta\phi)$; the naive count $Z-6$ is shown for contrast and does not collapse (more than half its points fall at $\Delta Z<0$).
(\textbf{C}) Count of quasi-zero modes $N_{\rm zero}$ versus the measured unconstrained-bud number $N_u$, validating Eq.~\eqref{eqS:nzero}.
}
\label{figS:dos}
\end{figure}

Figure~\ref{figS:dos} reports the measured vibrational spectrum. Panel~(\textbf{A}) shows $D(\omega)$ at several packing fractions between $\phi_J$ and $\phi_2$, for a population with strong feedback, exhibiting the predicted draining of the low-frequency band as buds are completed. The $N_u$ unconstrained-bud modes sit at $\omega\approx0$, separated from the band by a wide spectral gap because the arrest prestress is small; we therefore omit them here and count them directly in panel~(\textbf{C}). In panel~(\textbf{A}) only the vibrational band is shown. The low-frequency weight drains and the band edge $\omega^\ast$ rises as the unconstrained buds are completed. Panel~(\textbf{B}) collapses $\omega^\ast$, extracted from the inflection of the cumulative mode count $\int_0^\omega D(\omega')\,\dd\omega'$ (the onset of the extended band estimated as the RMS of its lowest few modes), onto a single line when plotted against the modified excess coordination $\Delta Z=\Delta Z_0+2(u_J-u)$ across the full $(P_0,\Delta\phi)$ set. For reference, we also plot the naive count $Z-6$, which does not collapse the data. Panel~(\textbf{C}) verifies Eq.~\eqref{eqS:nzero} by comparing the directly counted quasi-zero modes with the independently measured $N_u$. 
Together, these results confirm that the feedback-rigidified packings are marginal solids in the jamming universality class once the unconstrained-bud modes are identified~\cite{Wyart2005EPL,Brito2018,Ikeda2021}.

\section{Simulation protocol}\label{app:protocol}
The simulations follow the cell-based model introduced in Ref.~\cite{Gniewek2019}. Cells are represented as mother--bud dumbbells with repulsive harmonic lobe--lobe interactions. Growth proceeds by quasistatic increments of bud size under the local law
\begin{equation}
\gamma_i=\gamma_{0,i}\exp\!\left[-\frac{P_{\bud,i}}{P_0}\right],
\label{eqS:growthlaw}
\end{equation}
with complete overdamped mechanical relaxation after each increment.

The primary jamming point $\phi_J$ is identified as the first state with nonzero population pressure and a system-spanning force-bearing contact network. External bulk moduli are obtained by imposing a small isotropic areal strain on a mechanically arrested configuration and measuring the pressure response. Shear moduli are obtained analogously from small simple-shear deformations of the same arrested configurations. These measurements probe the timescale window $\tau_{\rm mech}\ll\tau_{\rm probe}\ll\tau_{\rm grow}$, where $\tau_{\rm mech}$ is the overdamped mechanical relaxation time, $\tau_{\rm probe}$ is the duration of the strain step, and $\tau_{\rm grow}$ is the growth time. 
In this regime it is assumed that the packing responds as a fixed contact network.
If instead $\tau_{\rm probe}\gtrsim\tau_{\rm grow}$, growth can create new contacts or reorganize the contact network during the deformation process, and the growth couples with the mechanical response of the packing.

\end{document}